\documentclass[12pt,a4paper,twoside]{scrartcl}
\usepackage{graphicx}
\usepackage{amsmath}
\usepackage[figuresright]{rotating}
\usepackage{fancyhdr}
\usepackage[width=16cm]{geometry} 
\newcommand{\TheAuthor}{}
\renewcommand{\TheAuthor}{S. Wycech}
\def\skiplinehalf{\medskip\\}

\def\supit#1{\raisebox{0.8ex}{\small\it #1}\hspace{0.05em}}  
\title{The physical interest in $K$$d$ 
and $\bar{p}$$d$ atoms }
\author{S.~Wycech\footnote{Invited talk to the International conference on Exotic Atoms, February 21-25, 2005, Vienna, Austria, proceedings "Verlag der \"{o}sterreichischen Akademie der Wisssenschaften"} \supit{a}, B.~Loiseau \supit{b}
\skiplinehalf
\supit{a}Soltan Institute for Nuclear Studies, Warsaw, Poland  \\
\supit{b}LPNHE Groupe Th\'eorie, Universit\'e P. et M. Curie, 
Paris, France }
\date{}
\begin{document}
\maketitle
\vspace{-0.5cm}
\begin{abstract}
Exotic deuterium and helium are discussed. 
The  $S,P$ and $D$  levels of  $\bar{p}$   and $K^{-}$  atoms 
are calculated. Absorptive, subthreshold  $\bar{p}N$   
 amplitudes are extracted from experimental data and  compared to 
model calculations. The existence 
of a  quasi-bound state 
in the $\bar{p}N$ system is indicated. In the $K^{-}$ atoms 
some  effects of    $\Sigma(1385)$ resonance   are evaluated. 
\end{abstract}
\vspace{-0.3cm}
\section{Introduction}

The lightest hadronic atoms offer a chance to  
test hadron-nucleon  scattering amplitudes at and just below 
the thresholds. This energy region is of special interest 
when  quasi-bound states exist in the hadron-nucleon system. 
The two cases of current interest,   $\bar{p}N $ and 
$KN$  belong to this  category.

Several experiments with antiprotonic hydrogen, deuterium and helium
atoms have been performed~\cite{augnp99,got99,aug99,sch91}.  These
require an analysis which is different from the standard optical model
description of heavier atoms. In particular, the best fit optical
model parameters bear no clear relation to the low energy $\bar{p}N $
scattering parameters.  However, there is a good chance to find such a
relation in the light atoms and we attempt it in this paper.

In the $KN$ system, two strange baryonic resonances 
$\Lambda(1405)$ and $\Sigma(1385)$  
are well known. However, their coupling to 
the $K$  meson nucleon  state is not fully understood. 
One point of interest in this field is the old question of 
the nature of $\Lambda(1405)$
state interpreted as an elementary particle~\cite{dali}, a $KN$ 
quasi-bound state~\cite{amar} or a mixture of both. 
The $1S \quad Kd$ level shift is expected to shed some light on this question.
Our particular interest in this system is motivated by the recent 
discovery of the nuclear $KNNN$~\cite{suz04} state and possible role of 
the $\Sigma(1385)$, \cite{wyc05}. We look 
for the effect of this resonance  upon the "upper"  widths of $K$ atoms.  

Two results are presented  in this paper:  

$\bullet$ 
The atomic level  shifts are related to the zero energy  scattering 
amplitudes  of the orbital hadron   on the atomic nucleus. 
In light $\bar{p}$  atoms the threshold parameters 
(lengths, volumes)  may be extracted from the   $\bar{p}d $ 
, $\bar{p} ^3$He, $\bar{p} ^4$He data. 
Next, a simple formula  is used which  
expresses these   parameters in terms of  
the $\bar{p} N$  amplitudes  averaged  over some 
subthreshold  energy region.   
The latter: lengths - $a$ and  volumes- $b$ 
are  taken as  free parameters and  are extracted from the data.
Such a program can  be  performed for   $S , P$ states in 
$\bar{p} d$ and $ P,D$ states in $\bar{p}$He.  In these 
nuclei, the  nucleons are bound differently   and thus 
different  $\bar{p} N$  energies are  involved. 
Thus,  one obtains  $a(E), b(E)$  which indicate distinct 
energy dependence in the subthreshold region. A unique resolution 
is obtained in the case of Im $a(E)$ and Im $b(E)$. In this 
case additional data from  the $\bar{p}$ stopped 
in $d$  and He chambers~\cite{BIZ74,BAL89}  allow to disclose 
the isospin content of the  absorptive  amplitudes.    
Finally, Im $a(E), b(E)$  are compared to the results 
of an updated Paris potential model~\cite{par05}. A good 
understanding of the data is obtained. It indicates  
a  $\bar{p} N$ quasi-bound  state in a   $P$ wave.  

$\bullet$
In the $K^{-}$ atoms we calculate the chances to reach the 
1S state in deuteron  and estimate the uncertainties 
involved in the related levels. A point of our  special interest  
is the contribution from the  $\Sigma(1385)$   to the 
upper level widths in $K-d$ and $K$-He systems.
\vspace{-0.3cm}

\section{The relation of level shifts to scattering amplitudes} 

Experiments which  detect the X-rays emitted 
from hadronic atoms provide atomic levels 
shifted and widened by  nuclear interactions. 
For a given $n$-th state of angular  momentum $L$  
the full energy $ E_{nL}$ is shifted from the electromagnetic 
level $ \epsilon_{nL}$ by a complex  level shifts 
$\Delta E_{nL} = \delta E_{nL} - i\Gamma_{nL}/2  = E_{nL}-\epsilon_{nL}$ 
This shift may be related to the corresponding  hadron nucleus 
scattering parameter $ A_L$, arising in the effective range 
expansion.  The  relation comes via  an  
expansion in $ A_L/ B^L$, as the  
Bohr radius $ B$ is usually much larger than 
the lengths  characteristic for the  nuclear interactions. 
For the $S$ waves 
\begin{equation}
\label{C1} 
E_{nS}-\epsilon_{nS}= 
\frac{2\pi}{\mu}|\psi_{n}(0)|^2A_{0}(1- \lambda A_{0}/B) 
\end{equation}
is known as  Trueman formula~\cite{true}. 
The  electromagnetic energy $ \epsilon_{nS}$ 
is composed of the Bohr  energy with  corrections  for 
relativity and the deuteron electric polarisability.  
In the $1S$ state $\lambda =3.154$,$ B_{\bar{p}d}=43.2fm$, $ A \approx 1fm$  
and the second order term in Eq.({\ref{C1}) makes a few 
percent correction. Such corrections are negligible in higher 
angular momentum states and a simpler  relation, \cite{lamb},  
 is sufficient 
\begin{equation}
\label{C2} 
\Delta E_{nL}= \epsilon_{nL}^{o} \frac{4}{n} \Pi_{i=1}^{L}
 ( \frac{1}{i^2}-\frac{1}{n^2} )A_{L}/B^{2L+1}.  
\end{equation}
These formulas generate the values shown in  table 1.

\begin{table}[htb]
\begin{center}
\caption{
 Level shifts in antiprotonic deuterium, 
 [keV]  for $S$ and [eV]  for $P$ states. Third column  gives the extracted 
${\bar p}d$ scattering parameters and an absorptive length 
obtained from ${\bar p}d$ scattering~\cite{obe}.
}
\label{table:1}
\newcommand{\m}{\hphantom{$-$}}
\newcommand{\cc}[1]{\multicolumn{1}{c}{#1}}
\renewcommand{\tabcolsep}{2pc} 
\renewcommand{\arraystretch}{1.2} 
\hfill \\
\begin{tabular}{@{}rrr}
\hline
          level         & $ \delta E-i\Gamma/2 $     &   $ A_L [fm^{2L+1}]$          \\
$ 1S $                  & 1.05(25)-i0.55(37) \cite{aug99}  & 0.71(16)- i0.40(27)         \\
$ S $  &                                  &         - i0.62(7) \cite{obe} \\
\hline 
$2P  $                  & 243(26) -i245(15)  \cite{aug99}  & 3.15(33) - i3.17(19) \\
\hline
\end{tabular}\\[2pt]
\end{center}
\end{table}

\vspace{-0.3cm}

\subsection{A formula for the antiproton-deuteron scattering length}

In this section the  threshold  ${\bar p} d$   scattering 
parameters $A_L$ are calculated. Later this method is extended to 
the P and D parameters in  He. 
The fine structure  is calculable but at this moment 
a meaningful discussion  may be done only  
with a spin and isospin averaged  $A_L$. 
These are 
obtained by a summation of  the ${\bar p}d$ 
multiple scattering series into a  quasi-geometric series. 
For a full  explanation of the method  we refer 
to~\cite{etad,pbard}, it compares 
successfully with exact calculations~\cite{del00}. 
The  multiple scattering series  for  scattering  
on two nucleons labeled n,p  below  
the deuteron breakup is 
\begin{eqnarray}
\label{f2}
T & = & t_n+t_p+t_nG_0t_p+t_pG_0t_n+t_nG_0t_pG_0t_n+t_pG_0t_nG_0t_p \nonumber \\
& + & (t_p+t_n)G_{NN}(t_p+t_n)+(t_p+t_n)G_{NN}(t_p+t_n)G_{NN}
(t_p+t_n)+...
\end{eqnarray}
where $t_i$ are ${\bar p}N$  scattering matrices, $G_0$ is
the free three-body propagator and $G_{NN} = G_0 T_{NN} G_0$
is that part of 
the three-body propagator which contains the nucleon-nucleon
scattering matrix $T_{NN}$. 
The  deuteron scattering amplitude is determined by an average
\begin{equation}
\label{f3} 
 T_{{\bar p} d}(E,L) =
<\psi_d j_{{\bar p}}^L \mid T \mid \psi_d j_{{\bar p}}^L>
\end{equation}
where $\psi_d$  is the deuteron wave function and $j_{{\bar N}}^L$ 
 is the  Bessel function for an $L$-th  partial wave  
in the antiproton-deuteron system. The  ${\bar p} d$ 
scattering parameters are  obtained in the  limit 
$ p \rightarrow 0$ of the expression  
$ A_L = - T_{{\bar p} d}(E,L)/[ p^{2L} (2\pi)^2 m_{{\bar p} d}]$
where $p$ is the  relative  momentum and $m_{{\bar p} d}$ is 
the corresponding reduced mass.  
The normalization  is related to a similar  normalisation 
of the two body matrices.  For $S$ waves 
$ t = - a(E) / [(2\pi)^2 \mu_{{\bar N}N}]$ 
where $a(E)$ is the scattering length. For  $P$ waves one needs 
a gradient operator which acts on the relative ${\bar p}N $ 
coordinates  
$ t = - 3 b(E)/[(2\pi)^2 \mu_{{\bar N} N}]
 \nabla  \nabla  $   
where $b(E)$ is the scattering volume. 
These matrices correspond to the four basic antiproton  
amplitudes of interest 
\begin{equation}
\label{f2a} 
f_{{\bar p} N}  = a_N(E)  +  3 b_ N(E) {\bf q} {\bf q}'
\end{equation}
where $ N$ stands for the proton or neutron 
and ${\bf q},{\bf q}' $ are the CM momenta before and after 
the collision.

The leading,  single  collision    $ < t_p + t_n  >$ term
yields the 
impulse  ${\bar p}d$ scattering parameters   
\begin{equation}
\label{f11} 
A_L  = \frac{\mu_{{\bar p}d}}{\mu_{{\bar p}N}}  \frac{1 }{[(2L+1)!!]^2}
[( \bar{a}_{{\bar p}p}+\bar{a}_{{\bar p}n} )<(r/2)^{2L}>_d  + 
( \bar{b}_{{\bar p}p}^1 +\bar{b}_{{\bar p}n})(\alpha +\beta <(r/2)^{2L-2}>_d ]
\end{equation}
where $<(r/2)^{2L}>_d$ is the $2L$-th radial moment  of the deuteron 
expressed in terms of the relative coordinate $r$ 
and  lengthy numerical factors $\alpha , \beta$   come from the derivative P-wave term.
In the limit of heavy nucleons,  which is not used here,  these  numbers 
may be found in~\cite{pbard}.   
Energy averaged values of  $ a$ and $b$ arise in Eq.( \ref{f11})    
\begin{equation}
\label{f12} 
\bar{a}(E)  =\int a \left(-E_d -\frac{p^2}{2m_{N,{\bar N} N}})\right) \mid
\tilde{\phi}_d^L(p)\mid^2 d\vec{p},  
 \end{equation}
as a result of  the spectator nucleon  recoil. 
The extent of the involved energies is determined by 
the nucleon binding and 
the Bessel transforms of the deuteron wave function 
\begin{equation}
\label{f13} 
 \tilde{\phi}_d^L(p) = \int \psi_d(r)j_L(pr/2) r^2 dr  
\end{equation}
These  energies  cover some unphysical  subthreshold region. 
The relevant distributions given by  Eq.(\ref{f12}) peak   
around $-12$, $-7$  and $-5$ MeV for $ L = 0,1,2 $ correspondingly. 
For heavier nuclei   and stronger nucleon bindings    
the energies of interest are shifted  further away  from the threshold.
That gives the chance to study the energy dependence of $ \bar{a}(E)$ 
and $ \bar{b}(E)$. 
A partial sum  of the series (\ref{f2}) 
is obtained by the first order geometric approximation 
\begin{equation}
\label{f5} 
A^1_L =\frac{ A_L}{1-\Omega -\Sigma}
\end{equation}
with $ \Omega = <t_pG_0t_n+t_n G_0t_p>/ <t_p + t_n > $ and $ \Sigma =
<( t_p + t_n ) G_{NN}( t_p + t_n) >/<t_p + t_n > $.  This partial sum
is $ 1 \% $ accurate in $2P$ states while in the $1S$ states a next
step is needed to reach that precision.  We also use these formulas to
describe the $2P$ and $3D$ levels in antiprotonic He, expressing
$G_{NN}$ by a simple separable nucleon-core interaction .
\vspace{-0.3cm}

\section{ The results}

\subsection{An extraction of the ${\bar p}N $ subthreshold amplitudes}
The formalism of the previous  section is now  used to extract the 
averaged  absorptive lengths  Im $a(E)$ and absorptive volumes Im $b(E)$ 
from the $d$ and He data.   The solution is not fully determined. 
For each atom one has four parameters Im $a$, Im $b$, Re $a$, Re $b$ 
and three atomic data -  the lower shift and two level widths.  In 
two cases, $d$ and  $^4$He, these are supplemented by the $S$ wave 
Im $A_0$ lengths~\cite{obe}. The best fit for 
absorptive parameters is possible but the real parts cannot 
be determined. 
The results are summarised in fig.1 and  compared with the  
updated Paris model calculation~\cite{par05}. This semi-phenomenological 
 model is based upon all available scattering data. 
More precisely,  the quantities plotted are  the half off-shell  amplitudes
$a,b(q=0,E,q(E))$. Two findings are of interest. 

First, there is an enhancement of the $P$ wave absorptive amplitude
just below the threshold. Within the model it corresponds to a
quasi-bound $^{31}P_1$ state. Another type of threshold enhancement
was recently found by the BES collaboration in the final ${\bar p}p
\gamma $ states obtained in the J$/\psi$ decays~\cite{bai03}.  That
enhancement may be attributed to the $^{11}S_0$ quasi-bound (or
virtual) state which is also generated by the Paris model. However,
the statistical weight of this state is too small to produce a clear
signal in the atomic data. Second, there is an increase of the $S$
wave absorption down below the threshold. Both these effects are
fairly well understood in terms of the model, although the threshold
result should be improved.

\begin{figure}[h]
\begin{center}
\includegraphics[width=0.55\textwidth]{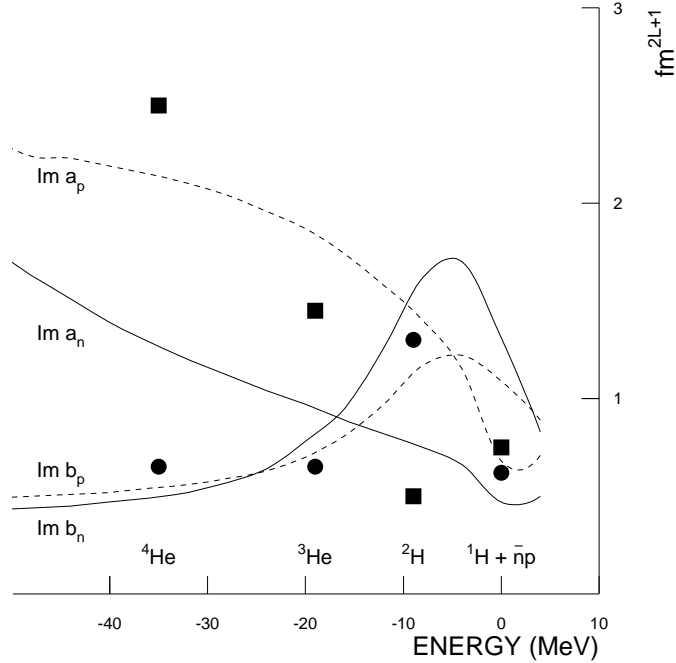}
\end{center}
\vspace{-0.3cm}
\caption{The absorptive parts of averaged 
subthreshold amplitudes calculated with Paris  model:  
 dotted lines - $\bar{p}p$, continuous lines - $\bar{p}n$ . 
The lengths are denoted as   $ a_p, a_n $  the   volumes 
$ b_p, b_n$. 
 The $b_p/2 + b_n/2$ should be 
compared   to the circles  which give  the average scattering  volumes 
extracted from $d$, $^3$He and $^4$He.  
In the same way  the extracted scattering lengths 
given by the squares  are to  be compared to $(a_p + a_n)/2$.
The experimental results at the threshold come from the 
$\bar{p}p$ atom 
and $\bar{n}p$ scattering experiment~\cite{mut88}.}
\end{figure}
\vspace{-0.3cm}

One can continue this analysis and cross  the threshold from below to 
study the absorption measurements of  ${\bar p} $ stopped in bubble 
chambers~\cite{BIZ74,BAL89}.
The ratios  of $S$ wave absorption obtained in this way and  given 
in table 2  are well reproduced by the Im $ a_n$, Im $a_p$ given in figure 1.

\begin{table}[htb]
\begin{center}
\caption{The  experimental antiproton capture ratios  
$R_{n/p}= \sigma (\bar{p}n)/ \sigma ( \bar{p}p)$
extracted   in states of flight  at very 
low energies.
}
\label{table:2}
\newcommand{\m}{\hphantom{$-$}}
\newcommand{\cc}[1]{\multicolumn{1}{c}{#1}}
\renewcommand{\tabcolsep}{2pc} 
\renewcommand{\arraystretch}{1.2} 
\hfill \\
\begin{tabular}{@{}lllll}
\hline
Element &  $R_{np}$  &   method  & state  &  reference    \\
\hline		       
d         &  0.81(3)  & chamber  & stopped   & \cite{BIZ74}    \\
$^{3}He$  &  0.47(4) & chamber  & stopped   & \cite{BAL89}    \\
$^{4}He$  &  0.48(3) & chamber  & stopped   & \cite{BAL89}    \\
\hline
\end{tabular}\\[2pt]
\end{center}
\end{table}

\vspace{-0.3cm}

\subsection{On the effects  of $\Lambda(1405)$ and $\Sigma(1385)$ 
on $Kd$ atomic levels }
The significance of $\Lambda(1405)$ in the optical potential studies 
of $K^-$ mesonic atoms is well known. It is the 
main factor that determines the mechanism of nuclear attraction. 
While in heavy nuclei the properties of $\Lambda(1405)$ may be 
changed by the  medium they   should be seen  clearly in the $1S$ 
level of the  $Kd$ atom. However, the knowledge of subthreshold 
$KN$ amplitudes is uncertain and detailed calculations depend on the way 
the $KN$ data is parameterised. This is exemplified by calculations 
from ref.~\cite{del00} given in table 3.
Different types of  amplitude parametrisations: 
constant lengths, K-matrix, separable potential fitted to the same data  
differ in the $1S$ level width by some $15 \%$. The experiment 
could be helpful in that respect. The scattering lengths in table 3 
seem  outdated now~\cite{zme05,iva05}  but the 
problem remains. It becomes more serious in  He due to  a rather  distant
extrapolation. The analysis performed in the last section 
for ${\bar p}$ may be repeated in the $K$ case when both 
the $Kd$ and $K$He data are available.  That seems possible 
even if the anomalous level shift in $^4$He~\cite{bird}  is confirmed.

\begin{table}[htb]
\begin{center}
\caption{ The $1S$   $Kd$  level shifts, comparison of several 
parametrisations of the same $KN$ data. Taken from 
ref.~\cite{del00}}
\label{table:3}
\newcommand{\m}{\hphantom{$-$}}
\newcommand{\cc}[1]{\multicolumn{1}{c}{#1}}
\renewcommand{\tabcolsep}{2pc} 
\renewcommand{\arraystretch}{1.2} 
\hfill \\
\begin{tabular}{@{}llll}
\hline
 & CSL &  K-matrix  & Separable    \\
\hline
 $a_{Kp}$ [fm]   &-0.623+i.763 & -0.663+i0.665 &-0.645 +i0.665    \\
$a_{Kn}$ [fm]   &0.322+i.748  & 0.264+i.570  & 0.32+i0.70                        \\
$\epsilon, \Gamma[KeV]$ &-0.50,~~1.02 & -0.49,~~0.89  &-0.50,~~0.98  \\
\hline
\end{tabular}\\[2pt]
\end{center}
\end{table}

\vspace{-0.3cm}

An interesting and open question is: what is the connection of the
 anomalous 2P shift in $^4$He to the P wave resonance
 $\Sigma(1385)$. We believe it is strong, \cite{wyc05}, but proper
 calculations are hard.  At this moment we look for effects of
 $\Sigma(1385)$ in upper atomic levels. These, perhaps, may be
 detected by the cascade intensity balance. The calculations based on
 the $S$ wave amplitudes from ref. \cite{amar} and SU(3) coupling to
 the $P$ wave $\Sigma(1385)$ \cite{wyc05} are presented in table 4.
 These are not very optimistic. The chance to observe the $2P$
 $\rightarrow$ $1S$ transition in deuterium is low. The $\Sigma(1385)$
 contributions to the $2P$, $3D$ widths are sizable but these widths would
 be difficult to extract from the cascade balance. The $3D$ level
 width in He, strongly affected by $\Sigma(1385)$, seems a bit easier
 to detect.

\begin{table}[h]
\begin{center}
\caption{The upper widths in $Kd$ and $K$$^4$He 
levels }
\label{table:4}
\newcommand{\m}{\hphantom{$-$}}
\newcommand{\cc}[1]{\multicolumn{1}{c}{#1}}
\renewcommand{\arraystretch}{1.2} 
\hfill \\
\begin{tabular}{@{}llll}
\hline
    & $Kd$, $\Gamma(2P)$ [meV]   & $Kd$,  $\Gamma(3D)$ [$\mu$eV] & $K$He,  $\Gamma(3D)$ [meV]     \\
\hline
Absorptive, S wave   & 26 & $ 10*10^{-2}$ & $2*10^{-2}$    \\
Absorptive, S + P    & 32 & $ 13*10^{-2}$ & $3.8 *10^{-2}$ \\
Radiative            & .32 &  32   & .57    \\
\hline 
\end{tabular}\\[2pt]
\end{center}
\end{table}
\vspace{-0.3cm}

\end{document}